\documentclass[a4paper]{article}

\usepackage{INTERSPEECH2020}

\title{EEG based Continuous Speech Recognition using Transformers}
\name{Gautam Krishna$^1$, Co Tran$^1$, Mason Carnahan, Ahmed H Tewfik}
\address{
  Brain Machine Interface Lab, The University of Texas at Austin\\
  $^1$equal author contribution}
\email{}

\begin{document}

\maketitle
\begin{abstract}
In this paper we investigate continuous speech recognition using electroencephalography (EEG) features using recently introduced end-to-end transformer based automatic speech recognition (ASR) model. Our results demonstrate that transformer based model demonstrate faster training compared to recurrent neural network (RNN) based sequence-to-sequence EEG models and better performance during inference time for smaller test set vocabulary but as we increase the vocabulary size, the performance of the RNN based models were better than transformer based model on a limited English vocabulary.
\end{abstract}

\section{Introduction}

Continuous speech recognition using non invasive brain signals or electroencephalography (EEG) signals is an emerging area of research where non invasive EEG signals recorded from the scalp of the subject is translated to text. EEG based continuous speech recognition technology  enables people with speaking disabilities or people who are not able to speak to have better technology accessibility. Current state-of-the-art voice assistant systems process mainly acoustic input features limiting technology accessibility for people with speaking disabilities or people with no ability to produce voice. In \cite{krishna2019speech} authors demonstrated deep learning based isolated speech recognition for recognizing five English vowels and four English words using only EEG features as input. In \cite{krishna20} authors demonstrated EEG based continuous speech recognition using state-of-the art end-to-end sequence-to-sequence recurrent neural network (RNN) based automatic speech recognition (ASR) models like CTC \cite{graves2014towards}, attention \cite{chorowski2015attention} on a limited English vocabulary consisting of 20 unique sentences. In \cite{krishna2019state} authors demonstrated EEG based continuous speech recognition using EEG signals recorded in parallel to spoken speech as well as using EEG signals recorded in parallel with listening utterances using different types of state-of-the-art ASR models on a limited English vocabulary consisting of 9 unique sentences. 

In \cite{vaswani2017attention} authors introduced a new type of sequence-to-sequence model named as transformer which can be applied to solve several sequence-to-sequence problems to get state-of-the-art performance on various tasks like machine translation \cite{vaswani2017attention}, language model \cite{devlin2018bert} and speech recognition \cite{li2019improving,dong2018speech}. Transformers use the concept of self attention, stacked layers of self attention, positional encoding \cite{vaswani2017attention} instead of recurrent networks like gated recurrent unit (GRU) \cite{chung2014empirical}, Long short-term memory (LSTM) \cite{hochreiter1997long} to learn sequence-to-sequence mapping. Transformers are faster to train compared to RNN models. To the best of our knowledge transformer models remains unexplored for EEG based continuous speech recognition task. In this paper we investigate EEG based continuous speech recognition used transformer model. We demonstrate our results on an English vocabulary consisting of 30 unique sentences during test time. Our results were better than RNN based model for smaller vocabulary size but as vocabulary size increase RNN based CTC model demonstrated better EEG recognition during test time \cite{krishna2019improving}.

\section{Transformer Speech Recognition model}
Figure 1 explains the architecture of the transformer ASR model used for mapping EEG features to text. The architecture we used in this work is very similar to the transformer model introduced by authors in \cite{vaswani2017attention}. The model at a higher level can be considered as an encoder-decoder model. The encoder model takes EEG features as input and applies non linear transformations to the input to produce a hidden representation which is fed into the decoder which again applies non linear transformations to produce text. Now we explain the encoder and decoder blocks in detail. The encoder is composed of stack of 8 identical layers. Similarly the decoder is also composed of stack of 8 identical layers. Each encoder layers consists of two sub layers namely multi head attention layer and fully connected layers. Each decoder layer is composed of three sub layers, where first two sub layers function same like the encoder sub layers whereas the third sub  layer in the decoder layer performs multi head attention on the output of encoder stack. Each sub layer in encoder and decoder is followed by layer normalization \cite{ba2016layer} and there exists residual connection around each of the sub layers \cite{vaswani2017attention}. We use masking in the multi head attention layer in the decoder to prevent it from depending on future positions \cite{vaswani2017attention}. 

The ${d_{model}}$ parameter was set to a value of 256. The ${d_{model}}$ parameter decides the output dimension of outputs of sub layers and embedding layers \cite{vaswani2017attention}. Both the embedding layers in the encoder and decoder block share the same set of weights. The details of self attention and multi head attention calculations are described in \cite{vaswani2017attention}. The parameter h was set to a value of 32. The parameter h refers to number of parallel attention layers or number of attention heads. 

The parameter ${d_{ff}}$ was set to a value of 1024. The parameter ${d_{ff}}$ refers to the number of hidden units in the fully connected sub layers in encoder and decoder layers. The parameters ${d_{k}}$ (queries, key vector dimension), ${d_{v}}$ (value vector dimension) \cite{vaswani2017attention} was set to 8. Basically multi head attention layer output is computed as concatenated outputs of individual attention heads multiplied by a weight matrix \cite{vaswani2017attention}. For implementing the positional encoding block shown in Figure 1 we used the same sine and cosine implementations used by authors in \cite{vaswani2017attention}. The key and value vectors from the final encoder layer is fed into the third multi head attention sub layer in the decoder layers. The multi head attention layer in the decoder takes query vector value from the layer beneath it. 

We used cross entropy as loss function for the model. After decoder block a dense layer is used to perform affine transformation and softmax activation is used to get output prediction probabilities.  
 During inference time we used a combination of beam search decoder and an external 4-gram language model, known as shallow fusion \cite{toshniwal2018comparison}. The label prediction process stops after the decoder predicts the end token. 

The model was trained for 120 epochs using adam \cite{kingma2014adam} optimizer and the batch size was set to 100. We used a word based model in this work. The model was predicting a word at every time step. 
All the scripts were written using keras and tensorflow 2.0 deep learning framework.

\begin{figure}[h]
\begin{center}
\includegraphics[height=8.5cm, width=\linewidth,trim={0.1cm 0.1cm 0.1cm 0.1cm}]{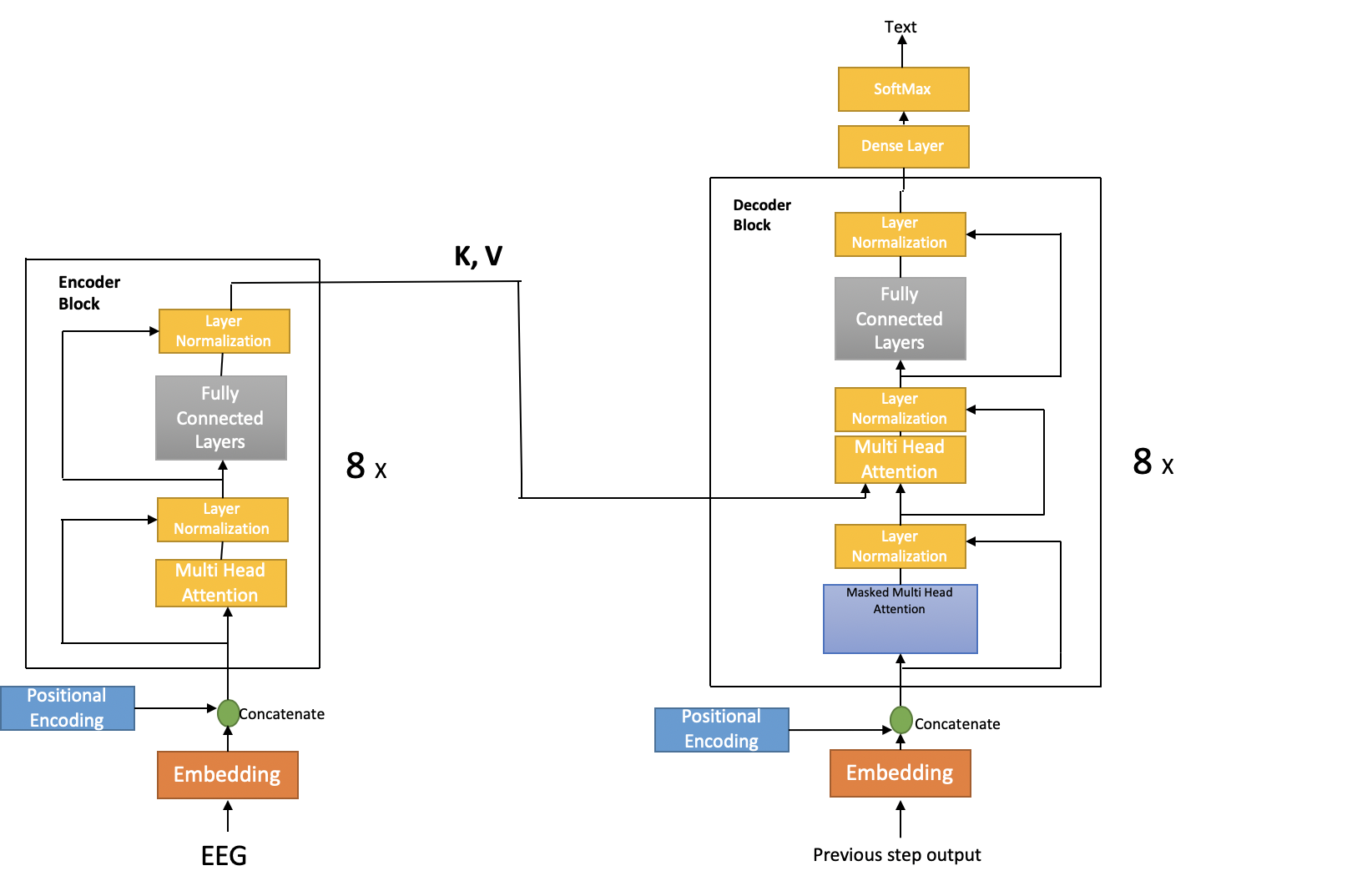}
\caption{Transformer ASR Model} 
\label{1vsall}
\end{center}
\end{figure}

\begin{figure}[h]
\begin{center}
\includegraphics[height=3cm,width=0.25\textwidth,trim={1cm 1cm 1cm 0.1cm},clip]{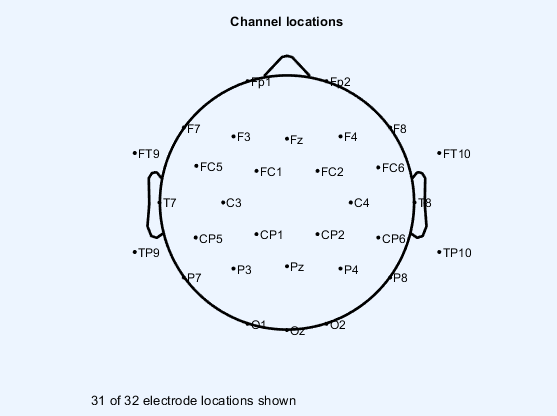}
\caption{EEG channel locations for the cap used in EEG experiments for collecting data} 
\label{1vsall}
\end{center}
\end{figure}

\section{Data Sets used for performing experiments}
We used data set A and data set B used by authors in \cite{krishna20} for this work. More details of the experiment design for collecting simultaneous speech and EEG data are covered in \cite{krishna20}. 

The authors in \cite{krishna20} used Brain product's ActiChamp EEG amplifier. Their EEG cap had 32 wet EEG electrodes including one electrode as ground as shown in Figure 2. They used EEGLab \cite{delorme2004eeglab} to obtain the EEG sensor location mapping. It is based on standard 10-20 EEG sensor placement method for 32 electrodes.

For each data set we used 80\% of the data as training set, remaining 10\% as validation set and rest 10\% as test set. The train-test split was done randomly. There was no overlap between training, testing and validation set. The way we splitted data in this work is different from method used by authors in \cite{krishna20}. 

\section{EEG feature extraction details}
We followed the same EEG preprocessing methods used by authors in \cite{krishna2019speech,krishna20}. 
EEG signals were sampled at 1000Hz and a fourth order IIR band pass filter with cut off frequencies 0.1Hz and 70Hz was applied. A notch filter with cut off frequency 60 Hz was used to remove the power line noise.
EEGlab's \cite{delorme2004eeglab} Independent component analysis (ICA) toolbox was used to remove other biological signal artifacts like electrocardiography (ECG), electromyography (EMG), electrooculography (EOG) etc from the EEG signals. %We used the ICA tool box from Brain Vision for this purpose.  
We extracted five statistical features for EEG, namely root mean square, zero crossing rate,moving window average,kurtosis and power spectral entropy \cite{krishna2019speech,krishna20}. So in total we extracted 31(channels) X 5 or 155 features for EEG signals.The EEG features were extracted at a sampling frequency of 100Hz for each EEG channel.

\begin{figure}[h]
\centering
\includegraphics[height=5cm, width=0.4
\textwidth,trim={0.1cm 0.1cm 0.1cm 0.1cm},clip]{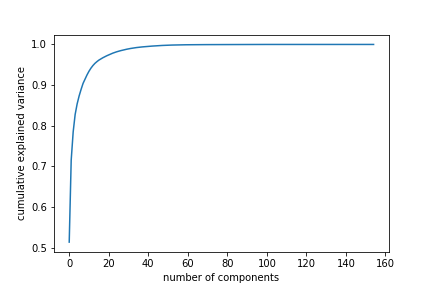}
\caption{Explained variance plot}
\label{1vsall}
\end{figure}

\section{EEG Feature Dimension Reduction Algorithm Details}
After extracting EEG features as explained in the previous section, we used Kernel Principle Component Analysis (KPCA) \cite{mika1999kernel} to denoise the EEG feature space as explained by authors in \cite{krishna20,krishna2019speech}. 
We reduced the 155 EEG features to a dimension of 30 by applying KPCA for both the data sets. We plotted cumulative explained variance versus number of components to identify the right feature dimension as shown in Figure 3. We used KPCA with polynomial kernel of degree 3 \cite{krishna2019speech,krishna20}. We then computed first and second order derivatives or delta, delta-delta coefficients of the 30 dimensional EEG features, thus the final EEG feature dimension was 90 (30 times 3). 

\begin{figure}[h]
\begin{center}
\includegraphics[height=6cm, width=\linewidth,trim={0.1cm 0.1cm 0.1cm 0.1cm}]{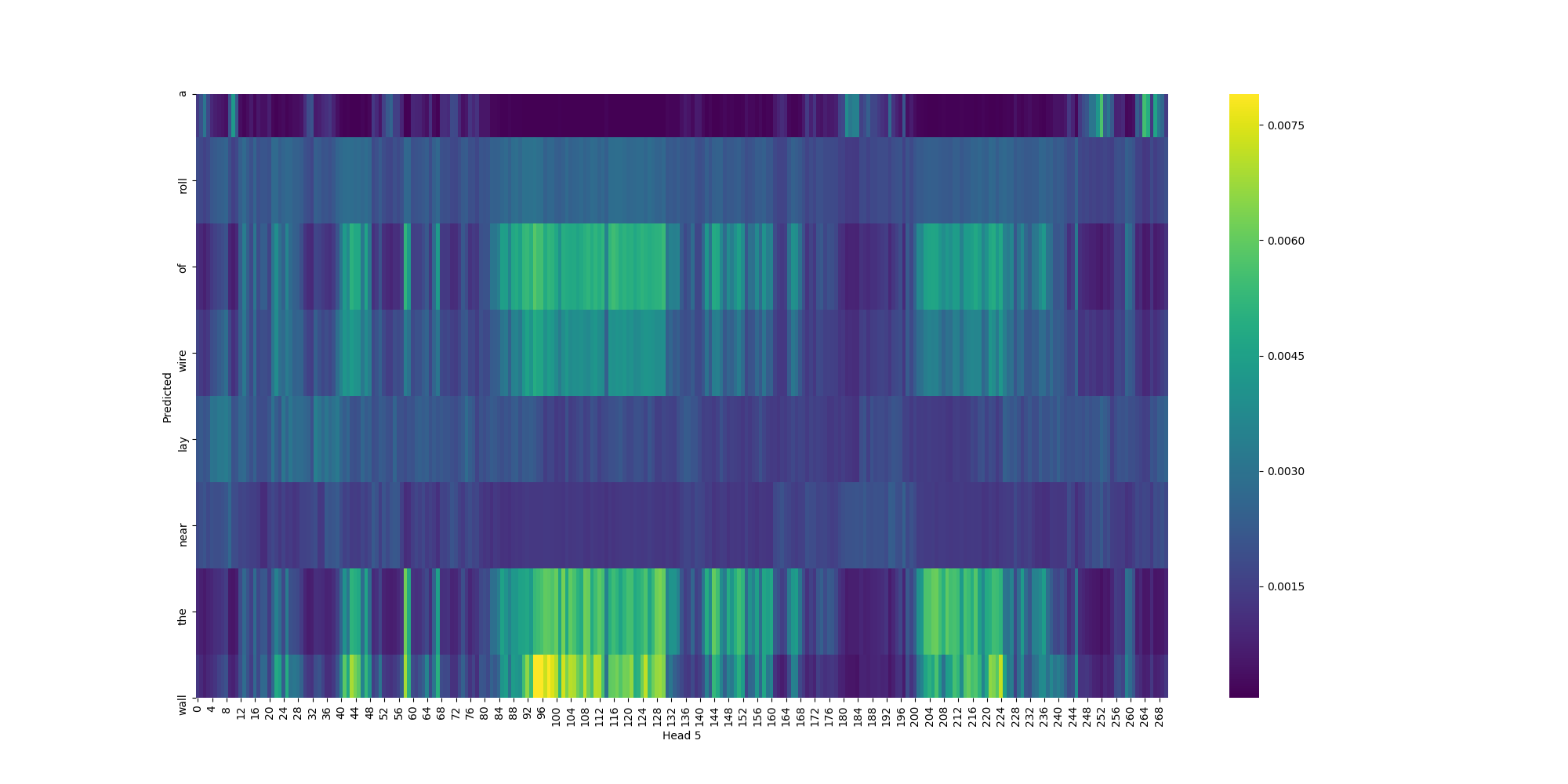}
\caption{visualization of attention weights for attention head 5} 
\label{1vsall}
\end{center}
\end{figure}

\begin{figure}[h]
\begin{center}
\includegraphics[height=6cm, width=\linewidth,trim={0.1cm 0.1cm 0.1cm 0.1cm}]{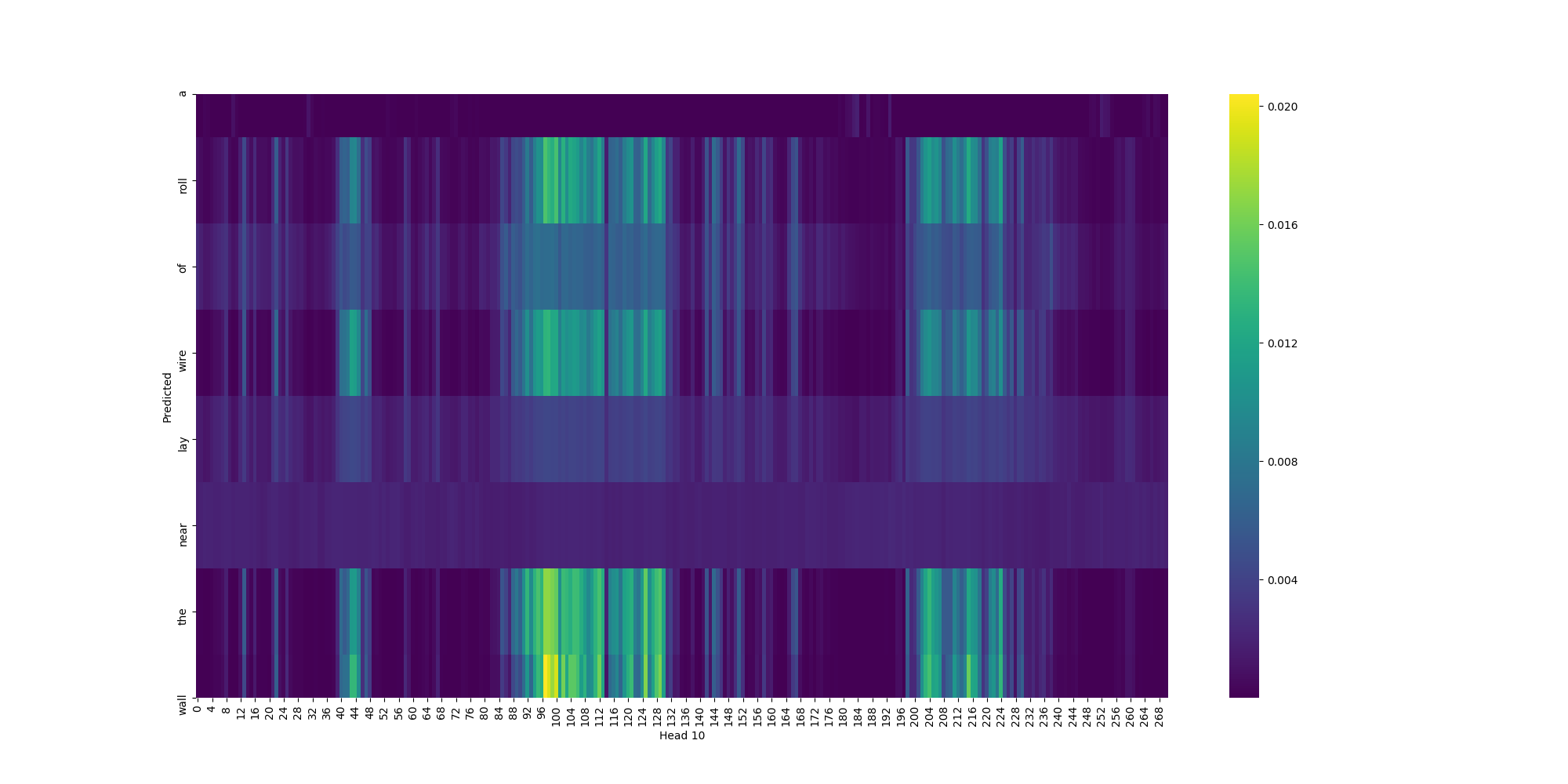}
\caption{visualization of attention weights for attention head 10} 
\label{1vsall}
\end{center}
\end{figure}

\begin{figure}[h]
\begin{center}
\includegraphics[height=6cm, width=\linewidth,trim={0.1cm 0.1cm 0.1cm 0.1cm}]{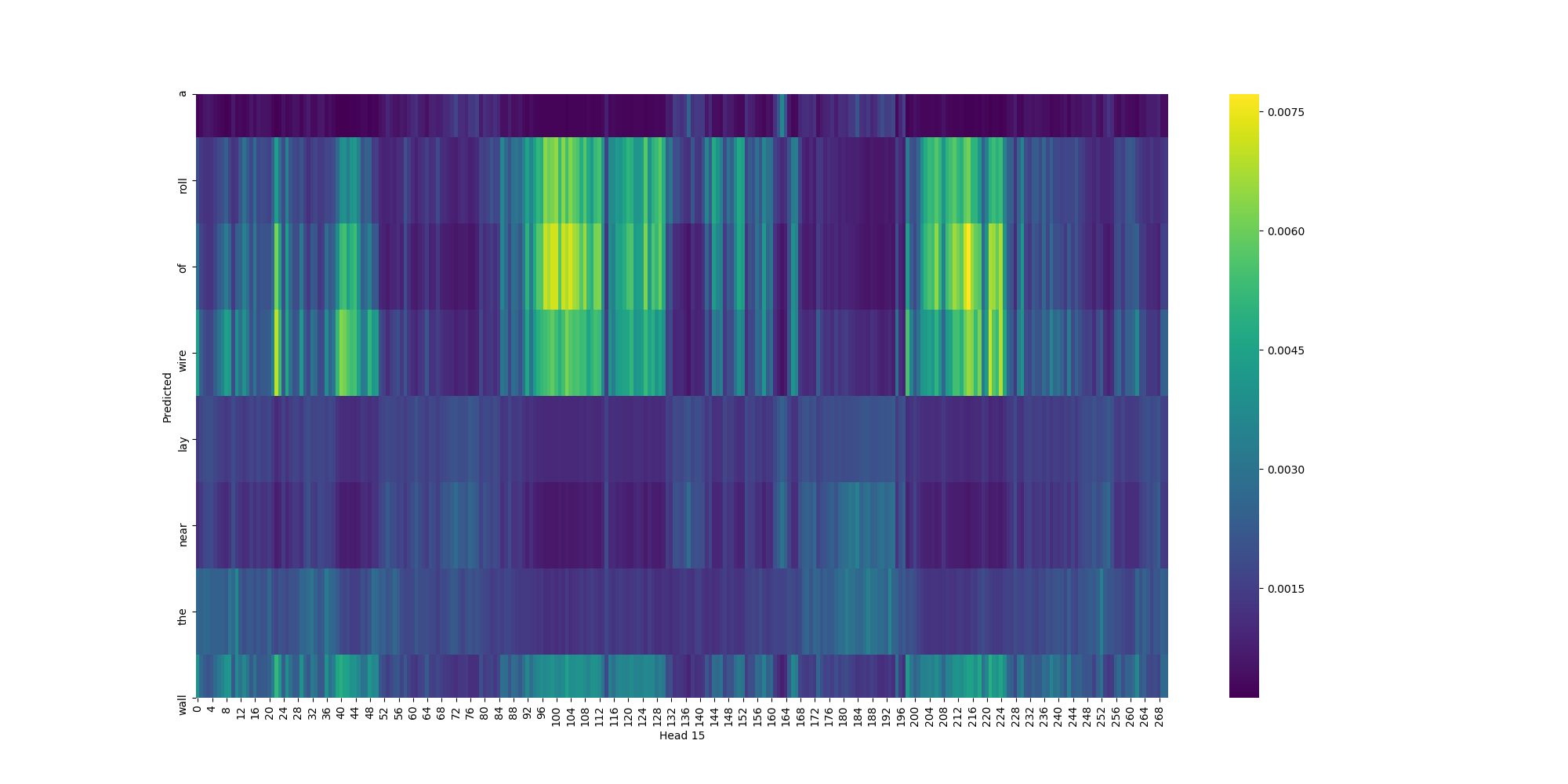}
\caption{visualization of attention weights for attention head 15} 
\label{1vsall}
\end{center}
\end{figure}

\section{Results}

We used word error rate (WER) as performance metric of the model during test time. Table 1 shows the results obtained during test time for Data set A. Table 2 shows the results obtained during test time for Data set B.

The average WER on test set is reported in both the tables. The transformer model demonstrated lower WER for smaller corpus size and WER went up as corpus size increase. Even-though transformer model demonstrated faster training and inference compared to RNN based end-to-end models like CTC, the transformer based results were poor compared to the results demonstrated in \cite{krishna2019improving} especially as corpus size increase.
Only when the test set consisted of \textbf{5 unique sentences} the transformer model demonstrated lower WER compared to the results demonstrated by authors in \cite{krishna2019improving} for rest of the test cases \cite{krishna2019improving} approach demonstrated lower WER. In \cite{krishna2019improving} authors used 20 \% of their total data as test set whereas in this work we used only 10 \% as test set to control over fitting. However the number of unique sentences and unique words contained in test set for each experiment in our work and for the work explained in \cite{krishna2019improving} were same. 

We also compared the training time and inference time for our transformer model with the CTC model explained in \cite{krishna2019improving} when both the models were trained for 120 epochs and with same batch size. For Data set A for the complete vocabulary consisting of 30 unique sentences, our transformer model took 0.017 minutes for training per example and 0.016 minutes for inference per example whereas the CTC model took 0.041 minutes for training per example and 0.0015 minutes for inference per example.

For Data set B for the complete vocabulary consisting of 30 unique sentences, our transformer model took 0.022 minutes for training per example and 0.015 minutes for inference per example whereas the CTC model took 0.043 minutes for training per example and 0.0022 minutes for inference per example.
The inference time also included the time taken to load the weights to the model during inference. We noticed the transformer model took more time to load the weights compared to the CTC model during inference as transformer model contained more number of parameters than the CTC model. Both transformer and CTC model used comparable amount of GPU memory for training. 

Interpretability of the visualization of the attention weights for various attention heads for EEG feature input still remains as a challenge. We tried plotting attention weights for various attention heads but we were not able to interpret the plots. Figures 4,5 and 6 shows some of the attention weight visualization plots. The attention weights basically learns the alignment between input EEG features and predictions (text) but since EEG signal is a complex signal it is not easy to interpret the attention weight plots directly. The attention weights magnitudes implies the importance given by the EEG features for predicting each word during inference time.

\begin{table}[!ht]
\centering
\begin{tabular}{|l|l|l|l|}
\hline
\textbf{\begin{tabular}[c]{@{}l@{}}Total \\ Number\\ of \\ Sentences\end{tabular}} & \textbf{\begin{tabular}[c]{@{}l@{}}Number\\  of\\ Unique\\ Sentences\\ Contained\end{tabular}} & \multicolumn{1}{c|}{\textbf{\begin{tabular}[c]{@{}c@{}}Number\\ of\\ Unique\\ words\\ Contained\end{tabular}}} & \textbf{\begin{tabular}[c]{@{}l@{}}EEG\\ WER\\ (\%)\end{tabular}} \\ \hline
15                                                                                 & 5                                                                                              & 29                                                                                                             & 67.7                                                              \\ \hline
30                                                                                 & 10                                                                                             & 59                                                                                                             & 83.95                                                             \\ \hline
45                                                                                 & 15                                                                                             & 84                                                                                                             & 88.85                                                             \\ \hline
60                                                                                 & 20                                                                                             & 106                                                                                                            & 91.04                                                             \\ \hline
75                                                                                 & 25                                                                                             & 132                                                                                                            & 91.15                                                             \\ \hline
90                                                                                 & 30                                                                                             & 153                                                                                                            & 93.95                                                             \\ \hline
\end{tabular}
\caption{WER on test set for Data set A}
\end{table}

\begin{table}[!ht]
\centering
\begin{tabular}{|l|l|l|l|}
\hline
\textbf{\begin{tabular}[c]{@{}l@{}}Total \\ Number\\ of \\ Sentences\end{tabular}} & \textbf{\begin{tabular}[c]{@{}l@{}}Number\\  of\\ Unique\\ Sentences\\ Contained\end{tabular}} & \multicolumn{1}{c|}{\textbf{\begin{tabular}[c]{@{}c@{}}Number\\ of\\ Unique\\ words\\ Contained\end{tabular}}} & \textbf{\begin{tabular}[c]{@{}l@{}}EEG\\ WER\\ (\%)\end{tabular}} \\ \hline
12                                                                                 & 5                                                                                              & 29                                                                                                             & 62.5                                                              \\ \hline
24                                                                                 & 10                                                                                             & 59                                                                                                             & 76.65                                                             \\ \hline
36                                                                                 & 15                                                                                             & 84                                                                                                             & 86.83                                                             \\ \hline
48                                                                                 & 20                                                                                             & 106                                                                                                            & 85.92                                                             \\ \hline
60                                                                                 & 25                                                                                             & 132                                                                                                            & 98.59                                                             \\ \hline
72                                                                                 & 30                                                                                             & 153                                                                                                            & 96.8                                                              \\ \hline
\end{tabular}
\caption{WER on test set for Data set B}
\end{table}

\section{Conclusion and Future work}
In this paper we explored EEG based continuous speech recognition using transformer sequence-to-sequence ASR  model.
The transformer model demonstrated lower WER or better results than RNN based model for smaller vocabulary size but as we increased the vocabulary size RNN based CTC model demonstrated better EEG recognition during test time \cite{krishna2019improving}. 
The interpretability of visualization of attention weights for various attention heads for EEG input still remains as an open problem. 

For future work we would like to build a larger Speech-EEG data base and validate our results on a larger English corpus. We would like to investigate whether our results can be improved by training the transformer model with a larger data set and also if attention weights interpretability can be improved by training the model with a larger data set. 

\section{Acknowledgement} 
We would like to thank Kerry Loader and Rezwanul Kabir from Dell, Austin, TX for donating us the GPU to train the models used in this work.

\bibliographystyle{IEEEtran}

\bibliography{mybib}

% \begin{thebibliography}{9}
% \bibitem[1]{Davis80-COP}
%   S.\ B.\ Davis and P.\ Mermelstein,
%   ``Comparison of parametric representation for monosyllabic word recognition in continuously spoken sentences,''
%   \textit{IEEE Transactions on Acoustics, Speech and Signal Processing}, vol.~28, no.~4, pp.~357--366, 1980.
% \bibitem[2]{Rabiner89-ATO}
%   L.\ R.\ Rabiner,
%   ``A tutorial on hidden Markov models and selected applications in speech recognition,''
%   \textit{Proceedings of the IEEE}, vol.~77, no.~2, pp.~257-286, 1989.
% \bibitem[3]{Hastie09-TEO}
%   T.\ Hastie, R.\ Tibshirani, and J.\ Friedman,
%   \textit{The Elements of Statistical Learning -- Data Mining, Inference, and Prediction}.
%   New York: Springer, 2009.
% \bibitem[4]{YourName17-XXX}
%   F.\ Lastname1, F.\ Lastname2, and F.\ Lastname3,
%   ``Title of your INTERSPEECH 2020 publication,''
%   in \textit{Interspeech 2020 -- 20\textsuperscript{th} Annual Conference of the International Speech Communication Association, September 15-19, Graz, Austria, Proceedings, Proceedings}, 2020, pp.~100--104.
% \end{thebibliography}

\end{document}